\documentclass[aps,prx,twocolumn,floatfix,showpacs,amssymb]{revtex4-1}
\usepackage[utf8]{inputenc}
\usepackage[english]{babel}
\usepackage{amsmath}
\usepackage{amsfonts}
\usepackage{amssymb}
\usepackage{graphicx}
\usepackage{float}
\usepackage{color}
\usepackage{braket}
\usepackage{mciteplus}
\newcommand{\be}{\begin{equation}}
\newcommand{\ee}{\end{equation}}
\newcommand{\ber}{\begin{eqnarray}}
\newcommand{\eer}{\end{eqnarray}}
\newcommand{\nn}{\nonumber}
\def\nablabold{\mbox{\boldmath $\nabla$}}

\newcommand{\pv}{{\bf p}}
\newcommand{\Ev}{{\bf E}}

\newcommand{\Dv}{{\bf D}}

\newcommand{\Iv}{{\bf I}}
\newcommand{\jv}{{\bf j}}
\newcommand{\rv}{{\bf r}}
\newcommand{\qv}{{\bf q}}
\newcommand{\kv}{{\bf k}}

\newcommand{\vv}{{\bf v}}

\def\rhov{\mbox{\boldmath $\rho$}}

\begin{document}

\title{Disorder-enabled hydrodynamics of charge and heat transport in monolayer graphene}

\author{Mohammad Zarenia$^1$,  Alessandro Principi$^2$, and Giovanni Vignale$^{1,3}$}
\affiliation{$^1$Department of Physics and Astronomy, University of Missouri, Columbia, Missouri 65211, USA\\
$^2$School of Physics, University of Manchester, Oxford Road, Manchester M13 9PL,
UK\\
$^3$Yale-NUS College, 16 College Ave West, 138527 Singapore}

\begin{abstract}
Hydrodynamic behavior in electronic systems is commonly 
accepted to be associated with extremely clean samples 
such that electron-electron collisions dominate and 
total momentum is conserved.  
Contrary to this, we show that in monolayer graphene
the presence of disorder is essential to enable an unconventional hydrodynamic regime which exists near 
the charge neutrality point and is characterized 
by a large enhancement of  the Wiedemann-Franz ratio.
 Although the enhancement becomes more pronounced
 with decreasing disorder, the very possibility of observing the 
effect depends crucially on the presence of disorder. 
We calculate the maximum extrinsic carrier density $n_c$ below 
which the effect becomes manifest, and show that $n_c$  
vanishes in the limit of zero disorder.  For $n>n_c$ we predict that the Wiedemann-Franz ratio
actually decreases with decreasing disorder. 
 We complete our analysis by presenting a transparent picture of the physical 
processes that are responsible for the crossover from conventional to 
disorder-enabled hydrodynamics.  Recent experiments on
 monolayer graphene are discussed and shown to be consistent with this picture.

\end{abstract}
%\pacs{81.05.ue , %graphene
%72.80.Vp , %Electronic transport in graphene
%}
\maketitle

\section{Introduction} Electric and thermal transport in electronic systems have long been described in terms of a single-particle picture~\cite{ashcroft} which emphasizes the role of collisions between electrons and impurities or phonons, while electron-electron interactions play a secondary role.
% sometimes as simple as a renormalization of the effective mass and other band structure parameters, in the spirit of the Landau theory of Fermi liquids.  
%Quantum mechanical corrections to the basic semiclassical theory of transport have also been intensively studied in the single-particle/Fermi liquid framework, exposing fascinating phenomena such as localization and metal-insulator transitions.  
It is only in the past two decades that advances in the fabrication of ultra clean samples have refocused the interest on collective hydrodynamic transport - a transport regime which is controlled by the nearly conserved quantities: number, momentum, and energy~\cite{Principi_prb_2016,Narozhny_prb_2015,Briskot_prb_2015,lucasReview}.
% that pertain to the electron liquid as a whole.

Under ordinary conditions (i.e., when the single-particle picture is valid) 
%the collective variables are rapidly degraded under the action of external defects (impurities, lattice vibrations), generically referred to as disorder.  For example, 
the total momentum of the electrons rapidly decays to the equilibrium value ($0$) as a result of electron-impurity collisions.  The motion of each electron, under these conditions, can be pictured as a random walk uncorrelated from that of the other electrons.
% (possibly with some quantum fuzziness) obeying the diffusion equation (if disorder is not too strong).  

The hydrodynamic regime provides a radical alternative to the single-particle scenario.  The electron liquid is in a state of local quasi equilibrium, which is established by frequent electron-electron collisions and is characterized by definite values of the collective variables, particle density, momentum density, and energy density.  Electron-electron collisions cannot change the global values of these conserved quantities (barring umklapp processes for momentum),  but can {\it redistribute} them over the system.  The momentum density, in particular, obeys a diffusion equation, with the viscosity playing the role of diffusion constant.  The motions of individual electrons, under these conditions, become correlated, since they must satisfy the constraints imposed by the conservation laws.

Of course, the hydrodynamic regime is not easily achieved for electrons in solids.  It is not sufficient that the system be very clean:  the temperature cannot be too high or too low because, in the first case, lattice vibrations destroy the conservation of momentum, and, in the second case, electron-electron collisions, hindered by the Pauli exclusion principle,  are not frequent enough to establish local quasi-equilibrium.  In spite of these difficulties, great progress has been made recently towards achieving this delicate balance in low-dimensional electronic materials, and several signatures of hydrodynamics,  such as an enhanced thermoelectric power near the charge neutrality point (CNP) of graphene\cite{ghahari}, viscosity-controlled vorticity~\cite{bandurin}, higher than ballistic conduction~\cite{kumar}, and the Gurzhi effect~\cite{Gurzhi}, have been experimentally observed.  
%(also mention applications of hydrodynamics to devices).

It might appear from the above description that disorder and hydrodynamics are mutually incompatible, since the former breaks the conservation of momentum, which is the basis for the latter.  In this paper we show, however, that this is not necessarily the case.
Specifically, we show that the thermoelectric transport properties of monolayer graphene near the charge neutrality point exhibit an intriguing  {\it disorder-enabled hydrodynamic regime}, namely a hydrodynamic regime which would not be observable in a perfectly clean system, but becomes observable in the presence of disorder within a range of doping densities that is proportional to the strength of disorder.  Indeed, such a regime has been observed in recent experiments \cite{crossno} and its most dramatic manifestation is a large enhancement of the Wiedemann-Franz ratio between the electric and the thermal resistivity in a narrow window of doping levels around the charge neutrality point.  

We remind the reader that the Wiedemann-Franz ratio, WF -- defined more precisely in the next section -- is a number of order 1 in the usual disorder-dominated single-particle transport regime. 
%($WF=\pi^2/3$ in the simplest model of static short-range disorder).  
This indicates that the dominant scattering mechanism affects equally electric and thermal transport.  In the hydrodynamic regime of an ordinary electron gas, on the other hand, one expects the electric resistivity to drop to values much smaller than the thermal resistivity because the conservation of momentum prevents Coulomb collisions from changing the particle current, while no such restriction exists for the thermal current: this situation results in a reduced value of the WF ratio \cite{vignale}.

{\it Doped} monolayer graphene is somewhat different from an ordinary electronic system in that the particle current does not coincide with the momentum and is therefore not automatically protected from decay as a result Coulomb collisions.  Nevertheless, even in this case the presence of a finite Fermi surface protects the particle current from decay due to Coulomb collisions and yields a reduced WF ratio \cite{Principi_conductivity,lucasWF,lucasKT,foster}.  

The situation changes in graphene near the charge neutrality point.  The protection arising from momentum conservation shifts from the particle current to the energy current which, in this regime, coincides with the thermal current and the total momentum~\cite{Fritz_2008,Muller_2008}.  
As a result, a large enhancement of the WF ratio has been predicted~\cite{Fritz_2008,Muller_2008}. However, we show that such
enhancement is observable only in a narrow window of doping densities -- a window that shrinks to zero as the system is made cleaner and cleaner.  Both features -- enhancement of WF ratio and shrinking window for its observation, are confirmed experimentally~\cite{crossno}.  This is, to the best of our knowledge, the first time that a disorder-enabled hydrodynamic regime has been demonstrated experimentally. 

While the phenomenon of the enhancement of the WF ratio near the CNP of graphene has attracted considerable theoretical interest in recent years (see in particular the early work{\bf s} in Refs. \cite{crossno,Fritz_2008,Muller_2008}, where the WF enhancement is discussed as an example of quantum critical behavior) most previous work on the subject has focused on detailed quantitative studies of the scattering mechanisms that control the size of the effect \cite{foster,svintsov2013,svintsov2018,shaffique}, including long-range inhomogeneities of the electron-hole density -- the so-called puddles~\cite{shaffique}.  In contrast to those previous studies, we take a minimalistic approach in which we first attempt to get a qualitative explanation in terms of electron-electron collisions only, then find that this approach greatly overestimates the effect, leading to mathematical singularities, and lastly introduce a minimal amount of disorder as a regularization tool to obtain finite and physically meaningful results.   We identify two distinct regimes of doping near the charge neutrality point: (i) a low-doping regime $n<n_c$ in which the WF ratio is much larger than the standard $\pi^2/3$ and increases with decreasing disorder, and (ii) a higher doping regime $n_c<n<n_{c2}$ in which the WF ratio is still larger than $\pi^2/3$ but now decreases with decreasing disorder. Both $n_c$ and $n_{c2}$ tend to zero in the limit of perfectly clean system, and we predict that measurements will inevitably fall into the second regime (ii) as the system is made less and less disordered at a given doping density.

Our analysis is based the simplest version of semiclassical transport theory, where the solution of the Boltzmann equation is parametrized in terms of two variables corresponding to the electric and the thermal current respectively. With the use of simple mathematics we display the crossover from disorder-enabled hydrodynamics to ordinary hydrodynamics, identify the doping densities at which the crossover occurs, and clarify the underlying physical mechanisms. 

%
% In what follows we present a detailed semi-analytical calculation of the thermoelectric coefficients of doped graphene monolayers.  We displays the cross-over from disorder-enabled to ordinary hydrodynamics, identify the disorder-controlled crossover density, and clarify the underlying physical mechanisms. 

\section{Definition and calculation of the thermoelectric resistivity matrix}

We work within the framework of quasi classical transport theory \cite{ashcroft}, where the state of the electrons is described by a homogeneous distribution function $f_{\kv,\gamma}$, where $\kv$ is the Bloch wave vector and $\gamma$ is the band index.  The deviation from equilibrium is  $\delta f_{\kv,\gamma}= f_{\kv,\gamma}-f^{(0)}_{\kv,\gamma}$, where $f^{(0)}_{\kv,\gamma}$ is the equilibrium distribution function at chemical potential $\mu$ and temperature $T$.   The quantities of interest are the electric current $\jv_e$ and the thermal current $\jv_q$, however, in order to homogenize the dimensions we will be working with the particle current $\jv_n = \jv_e/(-e)$ (where $-e$ is the charge of the electron) and the entropy current in units of $k_B$,  $\jv_s  =\beta \jv_q$, where $\beta=1/(k_BT)$.  These currents are related to the nonequilibrium distribution function by \cite{pines}
\be\label{EqJ}
\jv_n=\sum_{\kv,\gamma}\vv_{\kv,\gamma}\delta f_{\kv,\gamma},~~\jv_s=\sum_{\kv,\gamma}\beta\tilde{\epsilon}_{\kv,\gamma}\vv_{\kv,\gamma} \delta f_{\kv,\gamma}\,,
\ee
where $\tilde\epsilon_{\kv,\gamma}\equiv\epsilon_{\kv,\gamma}-\mu$, and $\epsilon_{\kv,\gamma}$ and  $\vv_{\kv,\gamma}$ are, respectively, the energy and the velocity of band $\gamma$ at wave vector $\kv$. 
The currents are connected to the electric field $\Ev$ and to the temperature gradient $\nablabold T$ by the thermoelectric resistivity matrix, $\rhov$, which we define as follows
 \be\label{EqJE}
 \left(\begin{array}{c}-e\Ev \\ -k_B\nabla T\end{array}\right)=\rhov\cdot\left(\begin{array}{c}\jv_n\\ \jv_s\end{array}\right)\,.
\ee
While this definition differs slightly from the conventional one \cite{ashcroft}, its main advantage is dimensional homogeneity, i.e., all the currents and all the components of the resistivity matrix have the same physical dimensions.  The elements of $\rhov$ are expressed in terms of three transport coefficients: the reduced electric resistivity $\bar \rho_{el}$ (this is the ordinary electric resistivity multiplied by $e^2$), the reduced thermal resistivity $\bar \rho_{th}$ (this is the usual thermal resistivity multiplied by $k_B^2 T$), and  the dimensionless Seebeck coefficient $\bar Q$ (this is the ordinary Seebeck coefficient expressed in units of $k_B/e$), in the following manner (see Ref.~\onlinecite{ashcroft})
\be\label{ResistivityMatrix}
\rhov=\left(\begin{array}{cc}\bar \rho_{el}+\bar Q^2\bar\rho_{th}& \bar Q \bar \rho_{th}\\ \bar Q \bar \rho_{th} & \bar\rho_{th}\end{array}\right)\,.
\ee
The determinant of this matrix is the product of the electric and thermal resistivities:
\be\label{Determinant}
{\rm det} \rhov = \bar\rho_{el} \bar\rho_{th}\,.
\ee
We also introduce the Wiedemann-Franz ratio, $WF$,  defined as follows
\be\label{WFRatio}
WF \equiv
\frac{\bar \rho_{el}}{\bar\rho_{th}} =
\frac{{\rm det} \rhov}{\bar\rho_{th}^2}\,,
\ee
where $WF=\pi^2/3$ when the Wiedemann-Franz law is satisfied in its standard form, e.g., for the classical model of a parabolic band with short-range disorder. 

In order to compute the transport coefficient we make the following 2-parameter Ansatz for the non equilibrium distribution function in the steady state:
\be\label{Ansatz}
\delta f_{\kv,\gamma}=f^\prime_{\kv,\gamma}\vv_{\kv,\gamma}\cdot \left[\pv_n+\beta\tilde\epsilon_{\kv,\gamma}\pv_s\right]\,,
\ee
where $f^\prime_{\kv,\gamma}$ denotes the derivative of the Fermi distribution with respect to energy,  $\pv_n$ and $\pv_s$ are momentum shifts associated with the particle and the entropy current, respectively:
\be\label{EqJ2}
\left(\begin{array}{c}\jv_n\\\jv_s\end{array}\right)=\Dv \cdot\left(
  \begin{array}{c}
    \pv_n\\
  \pv_s
  \end{array}
\right),
\ee
and $\Dv$ is a symmetric $2\times 2$ matrix of ``Drude weights", which are functions of $\mu$ and $T$.  These functions are calculated analytically and presented in Appendix A  together with their limiting forms for small and large values of the ratio $\bar \mu=\beta\mu$. 

The simple 2-parameter Ansatz of Eq.~(\ref{Ansatz}) suits our goal to keep the whole approach as simple as possible, while highlighting the essential physics of disorder-enabled hydrodynamics and achieving a good qualitative understanding of the main features observed in experiments~\cite{crossno}.
More complex 3-mode approximations~\cite{Narozhny_prb_2015,Briskot_prb_2015,Foster_prb_2009} have previously been employed in the derivation of hydrodynamic equations.  These are good to account for fine details of the physical picture, but do not change the basic features, which are nicely captured by our two-parameter Ansatz.
%We find that, using such approaches, the simplicity of the message of the present paper would be obscured. 
%Furthermore, the Ansatz~(\ref{Ansatz}) already enables us to achieve a good qualitative, if not quantitative, understanding of the main features observed in experiments~\cite{crossno}, as well as of the underlying mathematical structure.

We substitute Eq.~(\ref{Ansatz})  into the Boltzmann equation for the steady state response in the presence of fields $\Ev$ and $\nablabold T$ and require consistency between the assumed values of the currents and the ones that satisfy the Boltzmann equation,
\be\label{EqBZ}
-f^\prime_{\kv,\gamma}\vv_{\kv,\gamma}\cdot \left[e\Ev+\beta\tilde\epsilon_{\kv,\gamma}k_B\nablabold T\right]=I_{\kv,\gamma}.
\ee
The key input for this is the moments of the collision integral, $I_{\kv,\gamma}$, which to linear order in $\pv_n$ and $\pv_s$ are given by
\be\label{IMoments}
\left(\begin{array}{c}\sum_{\kv,\gamma}\vv_{\kv,\gamma}I_{\kv,\gamma} \\
\sum_{\kv,\gamma}\beta\tilde\epsilon_{\kv,\gamma}\vv_{\kv,\gamma}I_{\kv,\gamma}
\end{array}\right)=\Iv \cdot\left(
  \begin{array}{c}
    \pv_n\\
  \pv_s
  \end{array}
\right).
\ee
The $2\times 2$ matrix $\Iv$, which we refer to as ``collision kernel",  depends on the details of the microscopic scattering mechanism and is derived in Appendix B for Coulomb collisions and in Appendix C for our special model of elastic electron-impurity scattering.  Simple algebraic manipulations, using Eqs. (\ref{EqJE}), (\ref{Ansatz}) and (\ref{EqJ2}),  Eq. (\ref{EqBZ}) leads to the final expression for the thermoelectric resistivity matrix,
\be\label{BoltzmannResistivityMatrix}
\rhov = \Dv^{-1}\cdot\Iv\cdot\Dv^{-1}\,.
\ee
 This equation is the starting point of our analysis.  We note in passing that electron-electron interactions appear only in the collision kernel, $\Iv$.   The  ``Drude weights", $\Dv$, are completely determined by the clean noninteracting model. We expect that, in a more accurate  theory, they would be slightly renormalized by interactions.
 
\section{The limit of zero disorder}
\label{sect:zero_disorder}
If disorder is rigorously absent, then the collision integral is exclusively controlled by Coulomb collisions.  Regardless of any approximation we do in the treatment of these collisions, as long as Umklapp processes are neglected, the total momentum will be conserved, meaning that
\be\label{MomentumConservation}
\sum_{\kv,\gamma} \kv I_{\kv,\gamma}=0\,.
\ee
  This simple observation has huge consequences for the structure of the resistivity matrix.   First observe that in graphene $\epsilon_{\kv,\gamma}=\gamma \hbar v k$,  where $v$ is the Fermi velocity and $\gamma=1$ for the conduction band and $-1$ for the valence band, and $\vv_{\kv,\gamma}=\gamma v \hat \kv$, where $\hat\kv$ is the unit vector in the direction of $\kv$.
  This implies that the energy current is proportional to the momentum current:
  \be
  \sum_{\kv,\gamma} \epsilon_{\kv\gamma}\vv_{\kv,\gamma}\delta f_{\kv,\gamma}= v^2 \sum_{\kv,\gamma}\hbar \kv \delta f_{\kv,\gamma}\,,
  \ee
  which is a conserved quantity.   Substituting the expression of $\epsilon_{\kv,\gamma}$ and $\vv_{\kv,\gamma}$ in Eq.~(\ref{IMoments}), and making use of the conservation of momentum, Eq.~(\ref{MomentumConservation}), we get
  \be\label{IMoments2}
\left(\begin{array}{c}\sum_{\kv,\gamma}\vv_{\kv,\gamma}I_{\kv,\gamma} \\
-\bar \mu \sum_{\kv,\gamma}\vv_{\kv,\gamma}I_{\kv,\gamma}
\end{array}\right)=\Iv \cdot\left(
  \begin{array}{c}
    \pv_n\\
  \pv_s
  \end{array}
\right)\,
\ee
where $\bar \mu = \beta \mu$.  This being true for any value of $\pv_n$ and $\pv_s$  implies that  $\Iv$ has an eigenvector with eigenvalue $0$ -- a zero mode -- proportional to $\left(\begin{array}{c}\bar \mu\\1\end{array}\right)$.  Physically, this mode corresponds to a pure energy current, which is protected from decay by momentum conservation.

The presence of a zero mode implies that ${\rm det}\Iv=0$, and since the matrices of the Drude weights $\Dv$ are not singular (see Appendix A) we can immediately conclude that, {\it in the absence of disorder}
\be
{\rm det}\rhov=\bar\rho_{el}\bar\rho_{th}=0\,~~~({\rm clean~limit})\,.
\ee
This implies that either the electric resistivity or the thermal resistivity is zero for all doping levels and temperatures and, accordingly, the WF ratio can only be either zero or infinity.

%\be\label{ICoulomb}
%\Iv=\Iv_C=I_C\left(\begin{array}{c}1\\-\bar\mu\end{array}\right) \otimes \left(\begin{array}{cc}1\,, & -\bar\mu\end{array}\right)\,,
%\ee
%where the scalar quantity $I_C = \frac{{\rm Tr}\Iv_C}{1+\bar\mu^2}$ depends only on electron-electron interactions.
%Then the resistivity matrix takes the form
%\be\label{ICoulomb}
%\rhov=\rhov_C=I_C\left(\begin{array}{c}a\\b\end{array}\right) \otimes \left(\begin{array}{cc}a\,, & b\end{array}\right)\,,
%\ee
%where
%\be
%\left(\begin{array}{c}a\\b\end{array}\right) = \Dv^{-1}\cdot \left(\begin{array}{c}1\\-\bar\mu\end{array}\right)
%\ee 
%

Further analysis reveals that it is the electric resistivity that vanishes for any value of $n>0$.  
To see this, notice that the existence of the zero mode implies that the matrix $\Iv$ is completely determined by a single scalar function $I_C(\mu,T)$ and has the form
\be\label{ICoulomb}
\Iv=I_C\left(
\begin{array}{cc}
1
& -\bar{\mu}\\
-\bar{\mu} &
~~\bar{\mu}^2 \\
\end{array}
\right)\,.
\ee 
From this, the resistivity matrix is easily found to be
\be\label{RhoCoulomb}
\rhov=I_C\left(
\begin{array}{cc}
a^2
& -ab\\
-ab&
b^2\\
\end{array}
\right)\,,
\ee 
where
\be
a=\frac{D_{22}+\bar\mu D_{12}}{{\rm det} \Dv}\,,~~~~b=\frac{D_{21}+\bar\mu D_{11}}{{\rm det} \Dv}\,.
\ee
Limiting values of $a$ and $b$ for small and large values of $\bar \mu$ are given in Table I of Appendix A. In particular, near CNP ($\bar \mu\ll 1$),  we have
\be\label{a&b}
a(\bar\mu)\simeq\frac{(\pi \beta\hbar^2)}{\ln 4}\,,~~~~b(\bar\mu)\simeq\frac{2 \bar \mu (\pi \beta\hbar^2)}{9 \zeta(3)}\simeq 0\,.
\ee

Comparing this with Eq.~(\ref{ResistivityMatrix}) we immediately identify the transport coefficients as follows
\be
\bar\rho_{th}=I_C b^2\,,~~~\bar Q = -\frac{a}{b}\,,~~~\bar\rho_{el}=0\,,~~~(n>0)\,.
\ee

The only exception arises at the charge neutrality point where $\bar \mu=0$ and $b=0$ (see Appendix A) while $a$ remains finite. Thus, at the CNP the thermal resistivity vanishes, the electric resistivity acquires a finite value, and the Seebeck coefficient diverges:
\be\label{CNPCoefficients}
\bar\rho_{th}(0)=0\,,~~~\bar Q(0) = \infty\,,~~~\bar\rho_{el}(0)=\bar\rho_C=I_Ca^2\,.
\ee
(The $(0)$ argument emphasizes that this formulas refer to the CNP).
We emphasize that all of the above results follow from the conservation of momentum and are therefore valid only in the limit of zero disorder.   Some of these results are clearly unphysical: for example, the divergence of the Seebeck coefficient at CNP  and the discontinuous jump of the electric conductivity from zero to a finite value at $n=0$.  On the other hand, the electric resistivity at CNP, $I_C a^2$, and the thermal resistivity at finite doping levels,  $I_Cb^2$ are robust properties, in the sense that a small amount of disorder will affect their values as a small {\it perturbative} correction.  
\begin{figure}[t]
\includegraphics[width=9cm]{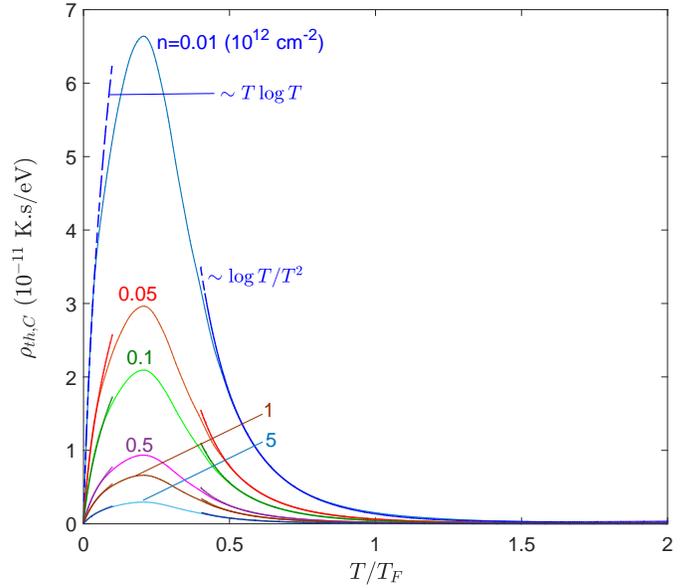}
\caption{Universal Coulomb thermal resistivity $\rho_{th,C}$  as a function of $T/T_F$ for different doping density as labeled.}
\label{fig1}
\end{figure} 

It is worth reflecting on the fact that the electric resistivity in the absence of disorder exhibits a discontinuity at the charge neutrality point.  At  this point the only carriers in the system are thermally excited electrons and holes in the conduction and valence bands respectively.  The two types of carries drift in opposite directions under the action of an electric field.  The physical origin of the finite resistivity is the transfer of momentum (also known as Coulomb drag \cite{CoulombDrag}) between the two types of carriers flowing in opposite directions.  Why does the resistivity plummet to zero as soon as extrinsic carriers are introduced in, say, the conduction band?  The answer is that the extrinsic carriers create a resistance-free channel of conduction which shunts the thermally activated electrons and holes.  Indeed, the extrinsic carriers act, in the absence of disorder, like the superfluid component of a superconductor:  they accelerate under the action of an electric field, while the thermally excited electrons and holes follow the displaced equilibrium distribution without experiencing any relative motion, hence without giving rise to Coulomb drag.  Thus, while our derivation has relied heavily on an approximate two-parameter solution of the Boltzmann equation, we believe that our qualitative conclusions has more general validity.

The remaining task for this section is to calculate the electron-electron collision kernel  $I_C$, which controls the resistivity matrix.  In Appendix B we make use of a standard approximation for the Coulomb collision integral (screened interaction plus Fermi golden rule) to find
\be\label{ICC}
I_C=-\frac{\beta }{4\pi }\sum_{\qv  }\int_{-\infty}^{\infty}d\omega\frac{|V(q)|^2 [(\Im\Pi_1)^2-\Im \Pi_0\Im \Pi_2]}{\sinh^2(\beta\hbar\omega/2)} \,,
\ee
where  the response functions $\Pi_n(q,\omega)$ ($n=0,1,2$) are defined as
\be\label{EqPn}
\Pi_n=4\sum_{\gamma,\gamma'}\sum_{\kv } \frac{F_{\kv ,\kv +\qv  }^{\gamma\gamma'}(v_{\kv ,\gamma} -v_{\kv +\qv  ,\gamma'})^n(f_{\kv ,\gamma} -f_{\kv +\qv  ,\gamma'})}{\epsilon_{\kv ,s} -\epsilon_{\kv +\qv  ,\gamma'}+\hbar\omega+i0^{+}}\,
\ee
$V(q)=2\pi e^2/\kappa(q+q_{\mathrm{TF}})$ is the screened Coulomb interaction with the Thomas-Fermi screening wave vector $q_{\mathrm{TF}}=4e^2k_F/(\kappa\hbar v )$.  The factor 4 accounts for the spin and valley degeneracy and $\kappa$ is the dielectric constant
of the substrate (within our calculations, we set $\kappa=4$ which is for h-BN substrate). 
The form factors $F_{\kv ,\kv +\qv  }^{\gamma\gamma'}=[1+\gamma\gamma'\cos(\theta_{\kv }-\theta_{\kv +\qv  })]/2$, where $\theta_\kv$ is the angle formed by the $\kv$ vector with the $x$-axis, come from the overlap of the wave functions at wave vectors $\kv$ and $\kv+\qv$.  
In Fig. \ref{fig1},  we plot the thermal resistivity as function of temperature (scaled with the Fermi temperature $T_F$) for different doping densities.  These curves are ``universal" in the sense that they depend only on the Coulomb interaction and on the doping density of the ideal clean model.  Our numerical results exhibit $\rho_{C,th}(T\to0)\sim T\ln T$ and 
$\rho_{C,th}(T\to\infty)\sim\ln T/T^2$ dependencies at low and high temperatures, respectively (see the fitting curves). 

The results for the electric resistivity, which is only finite at the CNP (first calculated by A. Kashuba in Ref.~\onlinecite{kashuba}), require the inclusion of a small amount of disorder and will be discussed in the next section. 

%and the  and the thermal conductivity at any doping level, are both expressed in terms of $I_C$ and the Drude weights and are plotted in Fig. 2 {\bf (Not true, check)}.  However, the WF ratio is zero for $n>0$ and infinity for $n=0$. In addition, the Seebeck coefficient, plotted in Fig. 3, is seen to have an unphysical singularity at $n=0$.  Evidently, the results of this section must be regularized if they are to make physical sense.
%

%%%%%%%%%%%%
%
\begin{figure}[]
\includegraphics[width=8.5cm]{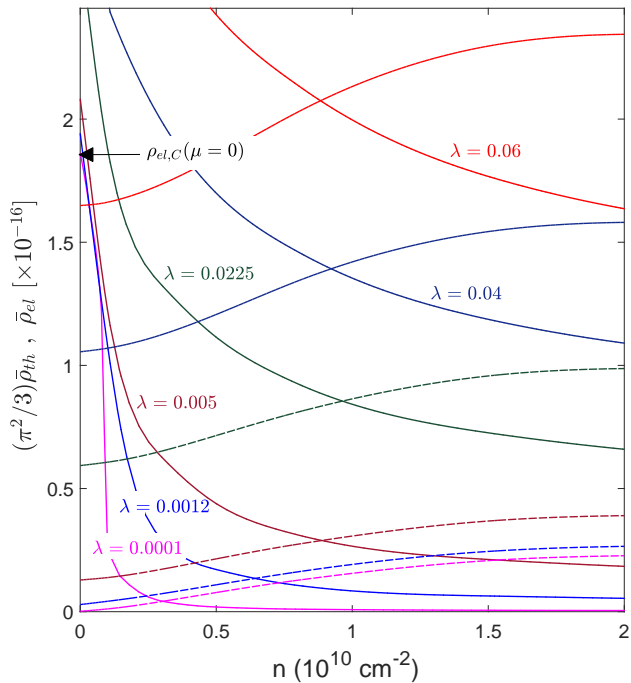}
\caption{Reduced electric (solid curves)  and thermal (dashed curves) resistivites as function of density for different disorder strength at $T=60$ K.}
\label{fig2}
\end{figure}
%\begin{figure}[t]
%\includegraphics[width=9cm]{fig4.eps}
%\caption{WF ratio as  a function of temperature for different carrier densities as labeled. The impurity strength is $u_0=0.1\hbar v $.}
%\end{figure}

\section{Disorder-enabled hydrodynamics}

Regularization of the results presented in the previous section is achieved  by including an infinitesimal amount of disorder, which breaks the conservation of momentum.  To see how this happens we add to the collision integral a small momentum-non-conserving part, i.e.,  we write
\be
\Iv=\Iv_C+ \Iv_D\,,
\ee
where the electron-electron collision kernel $\Iv_C$  (C for Coulomb) was defined in Eq.~(\ref{ICoulomb}), and the momentum-non-conserving kernel $\Iv_D$ (D for disorder) is assumed to be proportional to a dimensionless momentum relaxation rate $\lambda$, which we take to be $\ll 1$.  The precise form of $\Iv_D$ is not important for our purposes: the essential point is that it breaks the conservation of momentum.  However, for the sake of  illustration, we will later make use of a simple model of electrons and holes scattering from randomly distributed impurities of density $n_d$ with short-range potential $V_0\delta(\rv)$: for this model $\lambda = n_dV_0^2/(\hbar v)^2$, as detailed in Appendix C.  The resistivity matrix is the sum of two terms
\be
\rhov = \rhov_C+\rhov_D
\ee
where $\rhov_C$ is defined in Eq.~(\ref{RhoCoulomb}), 
and 
\ber
\rhov_D &=&  \Dv^{-1}\cdot  \Iv_D\cdot \Dv^{-1} \nn\\
&=& \left(\begin{array}{cc} \bar\rho_{el,D} +\bar Q_{D}^2\bar\rho_{th,D}&\bar Q_D \bar\rho_{th,D}\\ \bar Q_{D} \bar\rho_{th,D}&\bar\rho_{th,D}\end{array}\right)\,,
\eer 
where $\bar\rho_{el,D}$, $\bar\rho_{th,D}$, and $\bar Q_D$ are, respectively the electric resistivity, the thermal resistivity, and the Seebeck coefficient of the disordered system, {\it without including electron-electron collisions}.   In particular, $\bar\rho_{el,D}$, $\bar\rho_{th,D}$ are considered to be small quantities of order $\lambda\ll 1$.
\begin{figure*}[]
\includegraphics[width=12cm]{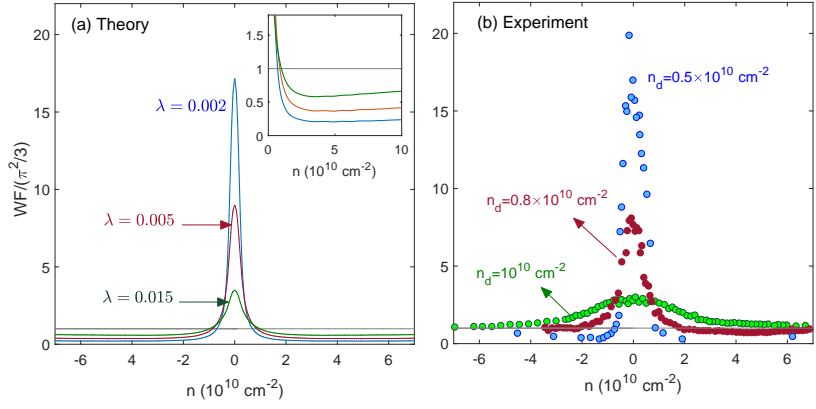}
\caption{(a) Theoretical WF ratio (scaled with $\pi^2/3$) as  a function of density at $T=60$ K for three different disorder strength as labeled. (b) Experimental WF ratio, taken from Ref. \cite{crossno}, for three graphene samples with different disorder density $n_d$. }
\end{figure*}\label{fig3}

From these formulas we extract our first important result, namely that the effect of disorder on the thermal conductivity is a small additive perturbation, $\bar \rho_{th,D}$:
\be\label{rhothermal}
\bar\rho_{th} = I_C b^2+\bar\rho_{th,D}\,.
\ee

To calculate the electric resistivity, we make use of the formula $\bar\rho_{el}=\rm{det}\rhov/\bar\rho_{th}$.  The determinant of $\rhov$ is most conveniently calculated in the basis of the (normalized) eigenvectors of the Coulomb resistivity matrix, $\rhov_C$, which are (see Eq.~(\ref{RhoCoulomb}))
\be\label{Eigenvector0}
|0\rangle\equiv 
\frac{1}{\sqrt{a^2+b^2}}
\left(\begin{array}{c}b \\ a\end{array}\right)
\ee
with eigenvalue $0$, and
\be\label{Eigenvector1}
|1\rangle\equiv 
\frac{1}{\sqrt{a^2+b^2}} 
\left(\begin{array}{c}a \\ -b\end{array}\right)
\ee
with eigenvalue $a^2+b^2$.  Notice that, at the charge neutrality point ($a$ finite, $b=0$), these eigenvectors reduce to $\left(\begin{array}{c}0 \\ 1\end{array}\right)$  and  $\left(\begin{array}{c}1 \\ 0\end{array}\right)$ respectively.
In the $|0\rangle, |1\rangle$ representation the resistivity matrix takes the form
\be
\rhov= \left(\begin{array}{cc} \langle0|\rhov_D|0\rangle &\langle0|\rhov_D|1\rangle\\ \langle1|\rhov_D|0\rangle&I_C(a^2+b^2) +\langle 1|\rhov_D|1\rangle\end{array}\right)\,,
\ee
and its determinant is readily seen to be
\be
{\rm det}\rhov =
I_C(a^2+b^2)  \langle0|\rhov_D|0\rangle+  \bar\rho_{el,D}\bar\rho_{th,D}\,,
\ee
from which we obtain
\be\label{rhoelectric0}
\bar\rho_{el} =\frac{I_C(a^2+b^2)\langle0|\rhov_D|0\rangle+\bar\rho_{el,D}\bar\rho_{th,D}}{I_C b^2+\bar\rho_{th,D}}\,.
\ee
We see that the effect of disorder on the electric resistivity is strongly non-additive except at the charge neutrality point, where $b=0$, $\langle0|\rhov_D|0\rangle = \bar\rho_{th,D}$, and we find the intuitive result
\be\label{RhoelLowDoping}
\bar \rho_{el}(0) = \bar\rho_C+\bar \rho_{el,D}(0)\,,
\ee
i.e., a small correction to the Coulomb drag resistivity.  The remarkable fact is that this formula remains essentially valid even at finite doping level (but we must have $\bar \mu \ll 1$) where the electric resistivity would be zero in the absence of disorder.  In other words, an infinitesimal amount of disorder causes the electric resistivity to jump from $0$ to $\bar\rho_C \equiv I_Ca^2$. 
The reduced thermal (dashed curves) and electric (solid curves) resistivities are plotted in Fig. \ref{fig2} as functions of density $n$ at $T=60$ K. For very low disorder  (e.g.,  $\lambda=0.0001$)  $\bar\rho_{th}$ vanishes and $\bar\rho_{el}$ approaches to $\bar\rho_C$ (indicated by an arrow)  at $n=0$. 

Now the question is: how low must the doping level be for the electric resistivity to remain close to $\bar \rho_C$?  To address this, we discard the second-order term  $\bar\rho_{el,D}\bar\rho_{th,D}\propto \lambda^2$  in the numerator of Eq.~(\ref{rhoelectric0}) and do some simple algebraic manipulations to rewrite
\be\label{rhoelectric}
\bar\rho_{el} \simeq \frac{\bar\rho_C\tilde \rho_D}{\bar\rho_C+\tilde\rho_D}\,,~~~~\tilde\rho_D = \bar \rho_{th,D}\frac{a^2}{b^2}\,,
\ee
%\be\label{rhoelectric}
%\bar\rho_{el} \simeq \frac{\bar\rho_C\bar \rho_{th,D}}{\bar\rho_C\frac{b^2}{a^2}+\bar\rho_{th,D}}\,,
%\ee
where we have used the fact that $b\ll a$ near the neutrality point. Indeed, from Eq.~(\ref{a&b}) we see that
\be\label{a/b}
\frac{a}{b} =\frac{2\pi^2}{3\bar \mu}
\ee
diverges at CNP (notice that this is the negative of the purely Coulombic Seebeck coefficient $\bar Q_C$).
\begin{figure}[]
\includegraphics[width=8cm]{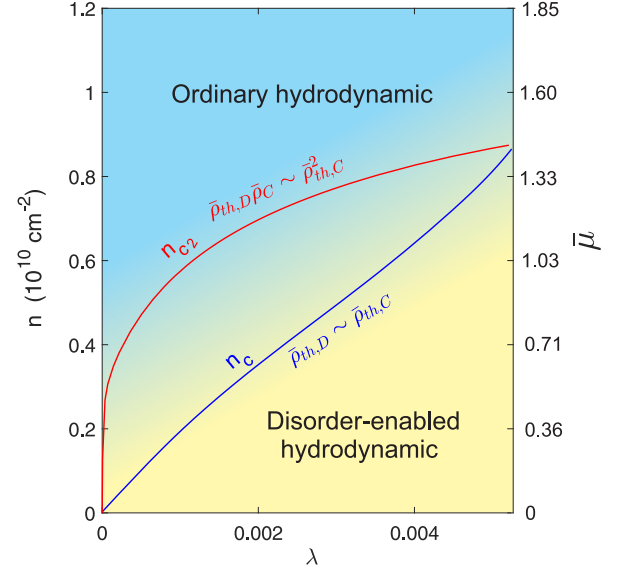}
\caption{The crossover densities $n_c$ and $n_{c2}$ (see text) as functions of the disorder strength $\lambda$ at $T=60$ K (The right y-axis shows the corresponding $\bar\mu$ values).}\label{fig4}
\end{figure}

This formula has an elegant physical interpretation in terms of two resistances in parallel.  Qualitatively, we see that the electric resistance remains essentially pinned to the Coulomb drag value, $\bar \rho_C$,  as long as  $\tilde\rho_D \gg \bar\rho_C$.  $\tilde\rho_D$ is proportional to disorder strength and therefore nominally small: however, it diverges as doping is reduced towards the charge neutrality point, because $b \to 0$.  This defines a crossover doping level, proportional to the strength of disorder,  below which the electric resistivity is controlled by Coulomb collisions.  We have called this the disorder-enabled hydrodynamic regime.  Of course, the above formula is valid for low doping levels.  In the opposite regime of $\bar \mu \to \infty$ the Coulomb kernel becomes negligible and the electric resistivity is controlled by disorder as usual.

The Wiedemann-Franz ratio provides a particularly incisive way to describe the crossover from disorder-enabled to ordinary hydrodynamics.  
%Noting that from Eq.~(\ref{rhothermal}) $\bar \rho_{th}$  can be rewritten as 
%\be
%\bar \rho_{th} = \frac{b^2}{a^2}\left(\bar\rho_C +\tilde\rho_{D}\right)\,.
%\ee
Upon combining Eq.~(\ref{rhothermal}) with Eq.~(\ref{rhoelectric}) (Eq.~(\ref{rhothermal})  being rewritten as $\bar\rho_{th}= \bar\rho_C\frac{b^2}{a^2}+\bar \rho_{th,D}$ -- see also Eq.~(\ref{CNPCoefficients}))  we obtain
\be\label{LorentzianSquared}
WF=\frac{\bar\rho_{el}}{\bar\rho_{th}} =\left(\frac{\Gamma}{\frac{b^2}{a^2}+\Gamma^2}\right)^2\,,~~~~\Gamma^2=\frac{\bar \rho_{th,D}}{\bar\rho_C}\,,
\ee
which is the {\it square} of a Lorentzian in the variable $b/a\simeq\frac{3\bar\mu}{2\pi^2}$. The Lorentzian has maximum value $\Gamma^{-1}$ and width at half maximum $\Gamma$.
This formula shows that the WF ratio at the neutrality point is greatly enhanced relative to its standard noninteracting value $\pi^2/3$:
\be
WF(0) =  \frac{1}{\Gamma^2}=\frac{\bar\rho_C}{\bar \rho_{th,D}}\,,
\ee
diverges as the strength of disorder, which $\bar\rho_{th,D}$ is proportional to, tends to zero.  This feature is clearly brought out by recent experimental measurements of WF in graphene near the neutrality point. In Fig. 3, we compare our theoretical $WF$ (Fig. 3(a))  for three different disorder strengths with the experimental results of Ref. \cite{crossno} measured in three samples with different disorder densities (Fig. 3(b)) . The numerical results confirm all the qualitative features deduced from the analytical work. 

Even more important from the present point of view is the fact that the enhancement of the WF persists in a doping window which we can (somewhat arbitrarily) define from the width of the Lorentzian at half maximum as $\frac{b^2}{a^2}<\frac{\bar \rho_{th,D}}{\bar\rho_C}$.  This coincides with the condition $\tilde\rho_D>\bar\rho_C$  deduced from the analysis of Eq.~(\ref{rhoelectric}) and defines a crossover doping density $n_c$ that scales linearly with the strength of disorder.  The doping region $n<n_c$ defines what we have dubbed disorder-enhanced hydrodynamic regime.  This regime is hydrodynamic because electron-electron collisions dominate over disorder and determine the enhancement of the WF ratio; it is disorder-enabled because the enhancement of $WF$ goes away  at any finite doping density when the strength of disorder goes to zero.   We emphasize that the $WF$ ratio remains much larger than $\pi^2/3$, in fact larger than $1/\Gamma^2 \gg \pi^2/3$, throughout the {\it disorder-enabled regime}.  
A more lenient crossover density $n_{c2}$ is obtained from the density for which $WF$ first drops below $\pi^2/3$ (see Fig. \ref{fig4}). We can say that the disorder-enabled hydrodynamic regime occurs for $n<n_{c2}$ {\it as long as  $n_c \ll n_{c2}$}.
When this is not the case, the $WF$ ratio becomes a rather structureless function of $n$ which does not differ much from the standard value $\pi^2/3$.  The electron liquid is no longer in the hydrodynamic regime.  

The two crossover lines $n_c$ an $n_{c2}$ are plotted in Fig.~\ref{fig4} and go to zero, respectively, as $\lambda^{1/2}$ and $\lambda^{1/4}$ in the clean limit $\lambda\to0$. 
\footnote{The $n_c$ line coincides with the $\bar\rho_{th,D}/\bar\rho_C\sim (b/a)^2$ and since $b/a(\bar\mu\to0)\sim\bar\mu^2$, see table I in Appendix A, we obtain $\lambda^{1/2}\propto\bar\mu$. The extrinsic density corresponding to the chemical potential $\mu$ is calculated using, 
$$
n=\frac{N_0}{\beta^2}\left[\mathrm{Li}_2(-e^{-\bar\mu})-\mathrm{Li}_2(-e^{\bar\mu})\right]
\approx
\frac{2\ln 2N_0}{\beta^2}\bar\mu
$$
where $N_0=2/\pi(\hbar v)^2$. With this we obtain $n_c(\bar\mu\to0)\sim\lambda^{1/2}$. The $n_{c2}$ crossover line, i.e. $WF\sim\pi^2/3$, corresponds to the  $ \bar\rho_{th,D}/\bar\rho_C\sim(b/a)^4$ condition and thus $n_{c2}\sim\lambda^{1/4}$ as $\lambda\to0$.}
 These lines divide the hydrodynamic regime into two regions: in region  (i), $0<n<n_c$, the WF ratio increases with decreasing $\lambda$, while in region (ii), $n_c<n<n_{c2}$, the WF ratio decreases with decreasing $\lambda$.  These behaviors are direct consequences of the analytic form of the squared Lorentzian of Eq.~(\ref{LorentzianSquared}).  From Fig. 4 we see that the $\lambda \to 0$ limit falls entirely within regime (ii) as long as the doping density is finite.   This means that the WF ratio at a given (small) doping density will eventually begin to decrease as the system is made cleaner and cleaner.  This prediction is nicely confirmed by the experimental data shown in Fig. 3b.
On the basis of this, we predict that the enhancement of the WF ratio will become visible, at a given doping density, only when the disorder strength exceeds a certain threshold.  In particular, the enhancement will not be observable when the crossover density $n_c$ drops below the level of unavoidable puddle fluctuations~\cite{shaffique}.

%The second condition insures that we have indeed a strong peak separating the doping density for which the peak drops to half its maximum value from the density for which it drops below $\pi^2/3$.  For $n>n_{c2}$ in this regime the system enters the {\it ordinary} hydrodynamic regime, where $WF<\pi^2/3$~\cite{vignale}.  

%All these distinctions are meaningful only as long as disorder is sufficiently weak: it must be so weak that the condition $n_c \ll n_{c2}$ is satisfied.  

\begin{figure}[]
\includegraphics[width=9cm]{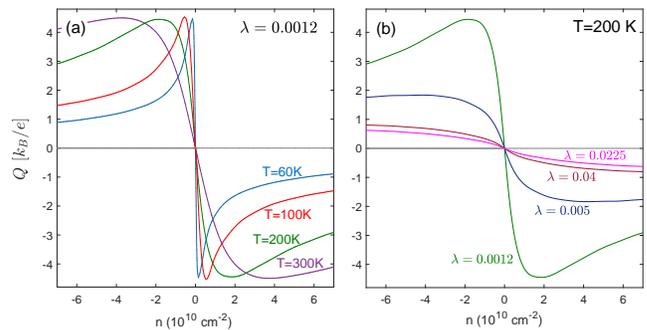}
\caption{Seebeck coefficient in graphene as a function of density for (a) different temperatures at a fixed disorder strength and for (b) different disorder strength at $T=200$ K. }
\label{fig5}
\end{figure}

As a final point, we notice that the inclusion of disorder also regularizes the Seebeck coefficient, which in the clean limit exhibits an unphysical divergence at the neutrality point.   In Fig. \ref{fig5} we see that the regularized Seebeck coefficient exhibits a large swing about the neutrality point, but goes to zero at the neutrality point as expected.  The swing region, in which the derivative of the $\bar Q$ vs doping density reverses its sign, is yet another incarnation of the disorder-enabled hydrodynamic regime.  Its width is defined by the same condition $|n|<n_c$ (or $|n|<n_{c2}$).

\section{Discussion, summary and conclusions}

The existence of an unusual hydrodynamic regime in graphene, characterized by a large enhancement of the Wiedemann-Franz ratio near the charge neutrality point,  has been recognized for a few years and has elicited considerable experimental and theoretical work.  The complex interplay of many scattering mechanism makes a quantitative study of the system quite difficult and may obscure the basic physics.  Rather than trying to compete with the many excellent works in the field, in this paper we have tried to reduce the phenomenon to its bare essentials, namely the conservation of momentum in Coulomb scattering  -- broken only by an infinitesimal amount of disorder which we introduce for the sake of regularization -- and the fact that momentum equals energy current in the regime of interest (whereas in usual electron liquids it is related to the particle current).  The quasi-conservation of the energy current implies a reduced thermal resistivity, but does not affect the electric resistivity, which is controlled by Coulomb drag between electrons and holes in different bands.  This is the basic fact underpinning the enhancement of the Wiedemann-Franz ratio.  Through the use of a bare-bones model we achieve great mathematical simplicity, traceability, and transparency in our description of the effect.  Having thus reduced the mathematical apparatus to a minimum we are able to gain fresh insight into the peculiar role played by disorder in establishing the new hydrodynamic regime.  We have realized that the new regime, which is latent in the structure of the Coulomb collision integral, would remain invisible without a small breaking of momentum conservation which creates the window in doping density in which the effect can be observed.  More precisely, we have been able to identify the doping density below which the enhancement of the WF ratio becomes significant and shown that this density tends to zero as the system becomes cleaner.  In particular, we predict that the effect will not be observable when the crossover density falls below the magnitude of the naturally occurring density fluctuations near the charge neutrality point  -- the so-called ``puddles" \cite{shaffique}. 
As a final point, we note that Umklapp scattering processes and other types of collisions, by breaking the conservation of momentum, could in principle offer alternative ways to regularize the results of Sect.~\ref{sect:zero_disorder}, at the price of introducing additional complexity. Some of these mechanism are discussed in excellent works that can be found in the literature~\cite{foster}. 
\begin{acknowledgements}
We are grateful to Shaffique Adam and Derek Ho Yew Hung for critical reading of our manuscript.
This work was supported by the U.S. Department of Energy (Office of Science) under grant
No. DE-FG02-05ER46203. MZ acknowledges hospitality and support from the Yale-NUS College and the Centre for Advanced 2D Materials in Singapore. 
\end{acknowledgements}

\bibliography{MZ}

%%%%%%%%%%%%%%
%

%%%%%%%%%%%%%%

\section*{Appendix A:  Drude weights}
\setcounter{equation}{0}
\renewcommand{\theequation}{A.\arabic{equation}}
In this section, we obtain explicit expressions for the $\Dv$-matrix elements, which we have dubbed ``Drude weights" and defined in Eq. (\ref{EqJ2}). Using Eqs. (\ref{EqJ}) and  (\ref{Ansatz}), the $\Dv$-matrix elements can be defined through the following equations
\be\label{EqD1}
\begin{split}
D_{11} &=\frac{g}{2}\sum_{\kv,\gamma}\vv_{\kv,\gamma}\cdot[f^\prime_{\kv,\gamma}\vv_{\kv,\gamma}],\\
D_{12} &=D_{21}=\frac{g}{2}\sum_{\kv,\gamma}\vv_{\kv,\gamma}\cdot[f^\prime_{\kv,\gamma}\vv_{\kv,\gamma}\beta\tilde{\epsilon}_{\kv,\gamma}],\\
D_{22} &=\frac{g}{2}\sum_{\kv,\gamma}\beta\tilde{\epsilon}_{\kv,\gamma}\vv_{\kv,\gamma}\cdot[f^\prime_{\kv,\gamma}\vv_{\kv,\gamma}\beta\tilde{\epsilon}_{\kv,\gamma}],
\end{split}
\ee
where the factor $g=4$ accounts for the spin and valley. $\tilde{\epsilon}_{\kv,\gamma}=\epsilon_{\kv,\gamma}-\mu$ with $\epsilon_{\kv,\gamma}=\gamma \hbar v  k$ ($\gamma=\pm 1$) are the linear bands of graphene and $\vv_{\kv,\gamma}=\gamma v \kv/k$ is the velocity vector at band $\gamma$. With these, Eqs. (\ref{EqD1}) lead to the following integrals 
\begin{widetext}
\be\label{eqD11}
\begin{split}
D_{11}&=\frac{2v_F^2}{(2\pi)^2}\sum_\gamma \int_{0}^{2\pi}d\theta\int_{0}^{\infty}kdk~ \frac{\kv }{\gamma k}. \frac{\kv }{\gamma k}\left(\frac{\partial f^0_{\kv ,\gamma }}{\partial \epsilon_{\kv ,\gamma }}\right)=
-\frac{1}{\pi\beta\hbar^2}\int_{0}^{\infty} dx~ x \left[\frac{ e^{x-\bar\mu}}{(1+e^{x-\bar\mu})^2}+\frac{ e^{-(x+\bar\mu)}}{(1+e^{-(x+\bar\mu)})^2}\right]\\
D_{12}&=D_{21}=\frac{2v_F^2\beta}{(2\pi)^2}\sum_\gamma \int_{0}^{2\pi}d\theta\int_{0}^{\infty}kdk~ \frac{\kv }{\gamma k}. \frac{\kv }{\gamma k}(\epsilon_{\kv ,\gamma }-\mu)\left(\frac{\partial f^0_{\kv ,\gamma }}{\partial \epsilon_{\kv ,\gamma }}\right)=
-\frac{1}{\pi\beta\hbar^2}\int_{0}^{\infty} dx~x \left[\frac{ (x-\bar\mu)e^{x-\bar\mu}}{(1+e^{x-\bar\mu})^2}-\frac{ (x+\bar\mu)e^{-(x+\bar\mu)}}{(1+e^{-(x+\bar\mu)})^2}\right]\\
D_{22}&=\frac{2v_F^2\beta^2}{(2\pi)^2}\sum_\gamma \int_{0}^{2\pi}d\theta\int_{0}^{\infty}kdk~ \frac{\kv }{\gamma k}. \frac{\kv }{\gamma k}(\epsilon_{\kv ,\gamma }-\mu)^2\left(\frac{\partial f^0_{\kv ,\gamma }}{\partial \epsilon_{\kv ,\gamma }}\right)=
-\frac{1}{\pi\beta\hbar^2}\int_{0}^{\infty} dx~x \left[\frac{ (x-\bar\mu)^2e^{x-\bar\mu}}{(1+e^{x-\bar\mu})^2}+\frac{ (x+\bar\mu)^2e^{-(x+\bar\mu)}}{(1+e^{-(x+\bar\mu)})^2}\right]
\end{split}
\ee
where we used the dimensionless variables $x=\beta\epsilon_+$ and $\bar\mu=\beta\mu$. The integrals (\ref{eqD11}) can be solved analytically resulting in the final expressions
\begin{table}[b]
\caption{Limiting behavior of the $\Dv$-matrix elements,  and the corresponding $a$ and $b$ defined in Eq. (\ref{RhoCoulomb}). $C=(\pi\beta\hbar^2)^{-1}$.}
\begin{tabular}{ r|lr}
  \hline\hline
   & $D_{11}\sim C\ln 4$ & $a\sim1/D_{11}$ \\
$\bar\mu\to0$   & $D_{12}\sim C\bar\mu\ln 4$ & $b\sim2\bar\mu/D_{22}$ \\ 
 &  $D_{22}\sim C9\zeta(3)$ & $b/a\sim[4\ln2/9\zeta(3)]\bar\mu$\\
  \hline \hline
  & $D_{11}\sim C\bar\mu$ & $a\sim2/D_{11}$ \\
 $\bar\mu\to\infty$   & $D_{12}\sim C\pi^2/3$& $b\sim1/D_{12}$\\ 
 &  $D_{22}\sim C\pi^2\bar\mu/3$ & $b/a\sim3\bar\mu/2\pi^2$ \\
  \hline\hline
\end{tabular}
\end{table}
\be\label{EqDf}
\begin{split}
D_{11}&=
-C\log[2(1+\cosh{\bar\mu})],\\
D_{12}&=D_{21}=
-C\left[-\frac{\pi^2}{3}- 2\bar\mu\log(1+e^{\bar\mu})-4\mathrm{Li}_2(-e^{\bar\mu})\right],\\
D_{22}&=-C\left[
-\frac{\bar\mu}{3} (\pi^2 - 6\bar\mu \log[1 + e^{\bar\mu}]) + 8 \bar\mu\mathrm{Li}_2(-e^{\bar\mu}) - 
 12 \mathrm{Li}_3(-e^{\bar\mu})
\right],
\end{split}
\ee
where $C=(\pi\beta\hbar^2)^{-1}$ and $\mathrm{Li}_n(z)=\sum_{k=1}^{\infty}z^k/k^n$ is the poly-logarithmic function. It would be helpful to look at the limits of Drude weights at $\bar\mu\to0$ (very low doping or high-$T$) and $\bar\mu\to\infty$ (high doping or low-$T$). The results are summarized in Table I. 
\section*{Appendix B: Electron-electron collision moments}
\setcounter{equation}{0}
\renewcommand{\theequation}{B.\arabic{equation}}
The electron-electron collision integral for the band $\gamma$ and at wave vector $\kv$, $I_{\kv ,\gamma}$ is given by, 
\begin{equation}
\begin{split}
I_{\kv,\gamma} &=-\sum_{\kv'}\sum_{\gamma',\eta,\eta'}\sum_{\qv }W(q)\big[ f_{\kv,\gamma }  (1-f_{\kv  -\qv,\gamma'})f_{\kv',\eta} (1-f_{\kv  '+\qv,\eta' })-f_{\kv -\qv,\gamma' }(1-f_{\kv,\gamma} )f_{\kv '+\qv,\eta' }(1-f_{\kv',\eta} )\big]\times\\
&\delta(\epsilon_{\kv,\gamma}  +\epsilon_{\kv',\eta} -\epsilon_{\kv  -\qv,\gamma' }-\epsilon_{\kv  +\qv ,\eta'})
\end{split}
\end{equation}
where the momentum conservation appears naturally when doing the second quantization
of Coulomb interaction in $\kv$-space, and the energy conservation stems from the Fermi golden
rule. $f_{\kv,\gamma}=f_{\kv,\gamma}^{(0)}+\delta f_{\kv,\gamma}$ is the non-equilibrium distribution function and $W(q)=(2\pi/\hbar)|V(q)|^2$
defines the collision probability where $V(\qv )$ is the statistic screened coulomb interaction. Inserting $\delta f_{\kv,\gamma}$ from Eq. (\ref{Ansatz}) and using

\begin{equation}
\delta(\epsilon_{\kv,\gamma}  +\epsilon_{\kv',\eta} -\epsilon_{\kv  -\qv,\gamma' }-\epsilon_{\kv  +\qv ,\eta'})=\hbar\int_{-\infty}^{\infty}d\omega
\delta(\epsilon_{\kv',\eta} -\epsilon_{\kv  '+\qv,\eta' }+\hbar\omega)\delta(\epsilon_{\kv,\gamma } -\epsilon_{\kv  -\qv,\gamma' }-\hbar\omega)
\end{equation}

and

\begin{equation}
f^0_{\kv} (1-f^0_{\kv\pm\qv })\delta(\epsilon_{\kv} -\epsilon_{\kv\pm\qv }\pm\hbar\omega)=
\frac{f^0_{\kv} -f^0_{\kv\pm\qv }}{\mp2e^{\pm\beta\hbar\omega/2}\sinh(\beta\hbar\omega/2)}\delta(\epsilon_{\kv} -\epsilon_{\kv\pm\qv }\pm\hbar\omega),
\end{equation}

we obtain

\begin{equation}\label{EqIk1}
\begin{split}
I_{\kv ,\gamma}=& \frac{2\pi \beta}{4}\sum_{\gamma',\eta,\eta'}\sum_{\kv^{\prime} }\sum_{\qv  }\int_{-\infty}^{\infty}d\omega\frac{|V(\qv  )|^2}{\sinh^2(\beta\hbar\omega/2)}\times\\
& F_{\kv ,\kv -\qv  }^{\gamma\gamma '}(f^0_{\kv ,\gamma } -f^0_{\kv -\qv  ,\gamma '})
\delta(\epsilon_{\kv ,\gamma } -\epsilon_{\kv -\qv  ,\gamma '}-\hbar\omega)
F_{\kv ',\kv '+\qv  }^{\eta\eta' }(f^0_{\kv ',\eta } -f^0_{\kv '+\qv  ,\eta '})\delta(\epsilon_{\kv ',\eta } -\epsilon_{\kv '+\qv  ,\eta '}+\hbar\omega)\times\\
&\left[\vv_{\kv ,\gamma }  -\vv_{\kv -\qv  ,\gamma '}+\vv_{\kv ',\eta } -\vv_{\kv '+\qv  ,\eta '}\right]\pv_n+ \beta\left[\tilde{\epsilon}_{\kv ,\gamma }\tilde{\epsilon}_{\kv ,\gamma }  -\vv_{\kv -\qv  ,\gamma '}\tilde{\epsilon}_{\kv -\qv  ,\gamma '}+\vv_{\kv ',\eta }\tilde{\epsilon}_{\kv ',\eta } -\vv_{\kv '+\qv  ,\eta '}\tilde{\epsilon}_{\kv '+\qv  ,\eta '}\right]\pv_s
\end{split}
\end{equation}
where $\{\gamma ,\gamma ',\eta ,\eta '\}=\pm 1$ denote the band index and $F_{\kv ,\kv -\qv  }^{\gamma\gamma '}$ is the form factor comes from the overlap of the wave functions at vector $\kv$ and $\kv+\qv$. For graphene, one can simply see that $\vv_{\kv ,\gamma }\tilde{\epsilon}_{\kv ,\gamma}=\kv-\mu\vv_{\kv ,\gamma}$. With this peculiar relation resulting from the linear dispersion of graphene, the last line in Eq. (\ref{EqIk1}) simplifies to
$\left[\vv_{\kv ,\gamma }  -\vv_{\kv -\qv  ,\gamma '}+\vv_{\kv ',\eta } -\vv_{\kv '+\qv  ,\eta '}\right](\pv_n-\bar\mu\pv_s)$.  Now the moments of the electron-electron collision integrals can be calculated using Eq. (\ref{IMoments2}),

\begin{equation}\label{EqIk2}
\sum_{\kv,\gamma}\vv_{\kv,\gamma}I_{\kv ,\gamma} = I_C (\pv_n-\bar\mu\pv_s),~~~
-\bar\mu\sum_{\kv,\gamma}\vv_{\kv,\gamma} I_{\kv ,\gamma}=I_C(-\bar\mu\pv_n+\bar{\mu}^2\pv_s)
\end{equation}
where the kernel $I_C$ is 
\begin{equation}\label{EqIC}
\begin{split}
I_C=&-\frac{\beta}{2\pi }\sum_{\qv  }\int_{-\infty}^{\infty}d\omega\frac{|V(\qv  )|^2}{\sinh^2(\beta\hbar\omega/2)} \sum_{\kv,\gamma ,\gamma '} v_{\kv ,\gamma }  F_{\kv ,\kv -\qv  }^{\gamma\gamma '}(f^0_{\kv ,\gamma } -f^0_{\kv -\qv  ,\gamma '})\delta(\epsilon_{\kv ,\gamma } -\epsilon_{\kv -\qv  ,\gamma '}-\hbar\omega)\times\\
&\sum_{\kv^{\prime},\eta ,\eta '}F_{\kv ',\kv '+\qv  }^{\eta\eta' }(f^0_{\kv ',\eta } -f^0_{\kv '+\qv  ,\eta '})\delta(\epsilon_{\kv ',\eta } -\epsilon_{\kv '+\qv  ,\eta '}+\hbar\omega)[\vv_{\kv ,\gamma }  -\vv_{\kv -\qv  ,\gamma '}+\vv_{\kv ',\eta } -\vv_{\kv '+\qv  ,\eta '}]. 
\end{split}
\end{equation}
We rewrite the first term in Eq. (\ref{EqIC}) as,
\begin{equation}
\begin{split}
&\sum_{\kv,\gamma ,\gamma '} v_{\kv ,\gamma } F_{\kv ,\kv -\qv  }^{\gamma\gamma '}(f^0_{\kv ,\gamma } -f^0_{\kv -\qv  ,\gamma '})\delta(\epsilon_{\kv ,\gamma } -\epsilon_{\kv -\qv  ,\gamma '}-\hbar\omega)=\\
&\frac{1}{2}\big[\sum_{\kv,\gamma ,\gamma '}v_{\kv ,\gamma }F_{\kv ,\kv -\qv  }^{\gamma\gamma '} (f^0_{\kv ,\gamma } -f^0_{\kv -\qv  ,\gamma '})\delta(\epsilon_{\kv ,\gamma } -\epsilon_{\kv  -\qv  ,\gamma '}-\hbar\omega)+\sum_{\kv,\gamma ,\gamma '}v_{\kv ,\gamma } F_{\kv ,\kv -\qv  }^{\gamma\gamma '}(f^0_{\kv ,\gamma } -f^0_{\kv -\qv  ,\gamma '})\delta(\epsilon_{\kv ,\gamma } -\epsilon_{\kv -\qv  ,\gamma '}-\hbar\omega)\big].
\end{split}
\end{equation}
Exchanging ($\kv \rightarrow \kv +\qv  $, $\gamma \to \gamma '$, $\gamma '\to \gamma $) in the first term and ($\kv \rightarrow {-\kv}$) in the second term we get
\begin{equation}
\begin{split}
&\sum_{\kv,\gamma ,\gamma '}v_{\kv ,\gamma }F_{\kv ,\kv -\qv  }^{\gamma\gamma '} (f^0_{\kv ,\gamma } -f^0_{\kv -\qv  ,\gamma '})\delta(\epsilon_{\kv ,\gamma } -\epsilon_{\kv -\qv  ,\gamma '}-\hbar\omega)
=\\
&\frac{1}{2}\big[\sum_{\kv,\gamma ,\gamma '} v_{\kv +\qv  ,\gamma '} F_{\kv ,\kv +\qv  }^{\gamma\gamma '}(f^0_{\kv +\qv  ,\gamma '} -f^0_{\kv ,\gamma })\delta(\epsilon_{\kv +\qv  ,\gamma '} -\epsilon_{\kv ,\gamma }-\hbar\omega)+\\
&\sum_{\kv,\gamma ,\gamma '} v_{-\kv ,\gamma } F_{\boldsymbol{-k},-\kv -\qv  }^{\gamma\gamma '}(f^0_{-\kv ,\gamma } -f^0_{-\kv -\qv  ,\gamma '})\delta(\epsilon_{-\kv ,\gamma } -\epsilon_{-\kv -\qv  ,\gamma }-\hbar\omega)\big]=\\
&\frac{1}{2}\sum_{\kv,\gamma ,\gamma '}F_{\kv ,\kv +\qv  }^{\gamma\gamma '}( v_{\kv ,\gamma }- v_{\kv +\qv  ,\gamma '}) (f^0_{\kv ,\gamma }-f^0_{\kv +\qv  ,\gamma '} )\delta(\epsilon_{\kv ,\gamma }-\epsilon_{\kv +\qv  ,\gamma '} +\hbar\omega)
\end{split}
\end{equation}
where we used  $\epsilon_{-\kv ,\gamma } =\epsilon_{\kv ,\gamma }$,  $f_{-\kv ,\gamma }  =f_{{\kv },\gamma }$, $F_{\kv ,\kv '}^{\gamma\gamma '}=F_{-\kv ,-\kv '}^{\gamma\gamma '}$, $v_{-\kv ,\gamma } =-v_{{\kv } }$ and exchanged $\omega\rightarrow-\omega$ which gives an overall  minus sign.  
With these transformations $I_C$ becomes
\begin{equation}\label{EqIC2}
\begin{split}
I_C &=-\frac{\beta}{4\pi}\sum_{\qv  }\int_{-\infty}^{\infty}d\omega\frac{|V(\qv  )|^2}{\sinh^2(\beta\hbar\omega/2)} \times\\
&\large[\sum_{\kv,\gamma ,\gamma '}F_{\kv ,\kv +\qv  }^{\gamma\gamma '}(v_{\kv ,\gamma }-v_{\boldsymbol{k+q},\gamma '})  (f^0_{\kv ,\gamma } -f^0_{\kv +\qv  ,\gamma '})\delta(\epsilon_{\kv ,\gamma }+\epsilon_{\kv +\qv  ,\gamma '}+\hbar\omega)\large]^2 +\\
&2[\sum_{\kv^\prime,\eta ,\eta '}F_{\kv ,\kv +\qv  }^{\gamma\gamma '}(f^0_{\kv ',\eta } -f^0_{\kv '+\qv  ,\eta '})\delta(\epsilon_{\kv ',\eta } -\epsilon_{\kv '+\qv  ,\eta '}+\hbar\omega)]\times\\
&[\sum_{\kv,\gamma ,\gamma '}F_{\kv ,\kv -\qv  }^{\gamma\gamma '}\vv_{\kv ,\gamma }(\vv_{\kv ,\gamma }  -\vv_{\kv -\qv  ,\gamma '})(f^0_{\kv ,\gamma } -f^0_{\kv -\qv  ,\gamma '})\delta(\epsilon_{\kv ,\gamma } -\epsilon_{\kv -\qv  ,\gamma '}-\hbar\omega)]
\end{split}
\end{equation}

Similarly,  we rewrite the last term as
\begin{equation}\label{EqIC3}
\begin{split}
&\sum_{\kv,\gamma ,\gamma '}F_{\kv ,\kv -\qv  }^{\gamma\gamma '}v_{\kv ,\gamma }(v_{\kv ,\gamma }    -v_{\kv -\qv  ,\gamma '}) (f^0_{\kv ,\gamma } -f^0_{\kv -\qv  ,\gamma '})\delta(\epsilon_{\kv ,\gamma } -\epsilon_{\kv -\qv  ,\gamma '}-\hbar\omega)=\\
&\frac{1}{2}\sum_{\kv,\gamma ,\gamma '}F_{\kv ,\kv +\qv  }^{\gamma\gamma '} v_{\kv +\qv  ,\gamma '}(v_{\kv +\qv  ,\gamma '}-v_{\kv ,\gamma }) (f^0_{\kv +\qv  ,\gamma '} -f^0_{\kv ,\gamma })\delta(\epsilon_{\kv +\qv  ,\gamma '} -\epsilon_{\kv ,\gamma }-\hbar\omega)+\\
&\frac{1}{2}\sum_{\kv,\gamma ,\gamma '} -F_{-\kv ,-\kv -\qv  }^{\gamma\gamma '}v_{\kv ,\gamma }(-v_{\kv ,\gamma } +v_{\kv +\qv  ,\gamma '}) (f^0_{\kv ,\gamma } -f^0_{\kv +\qv  ,\gamma '})\delta(\epsilon_{\kv ,\gamma } -\epsilon_{\kv +\qv  ,\gamma '}-\hbar\omega)=\\
&-\frac{1}{2}\left[\sum_{\kv,\gamma ,\gamma '} F_{\kv ,\kv +\qv  }^{\gamma\gamma '}(v_{\kv ,\gamma } -v_{\kv +\qv  ,\gamma '})^2(f^0_{\kv ,\gamma } -f^0_{\kv +\qv  ,\gamma '})\delta(\epsilon_{\kv ,\gamma } -\epsilon_{\kv +\qv  ,\gamma '}+\hbar\omega)\right]
\end{split}
\end{equation}
Using Eq. (\ref{EqIC3}), and rewriting the $\delta$ functions in terms of the imaginary parts of the response functions $\Pi_n$,  given in Eq. (\ref{EqPn}) in the main text, we finally obtain
\begin{equation}\label{EqICf}
I_C  =-\frac{\beta }{4\pi }\sum_{\qv  }\int_{-\infty}^{\infty}d\omega\frac{|V(\qv  )|^2}{\sinh^2(\beta\hbar\omega/2)} [(\Im\Pi_1)^2-\Im \Pi_0\Im \Pi_2]
\end{equation}

\section*{Appendix C: Disorder collision moments }
\setcounter{equation}{0}
\renewcommand{\theequation}{C.\arabic{equation}}
The non-momentum-conserving disorder collision integral is given by,
\begin{equation}
\begin{split}
\Iv_D(\kv ,\gamma)  =\frac{2\pi g}{\hbar}\sum_{\gamma '}\sum_{\boldsymbol{k'}}|U_{\kv -\boldsymbol{k'}}|^2 F_{\kv ,\boldsymbol{k'}}^{\gamma\gamma'}(f_{\kv ,\gamma } -f_{\kv ',\gamma '})\delta(\epsilon_{\kv ,\gamma }-\epsilon_{\kv ',\gamma '})
\end{split}
\end{equation}
where the factor $g=4$ accounts for the spin and valley degeneracy and $F_{\kv ,\boldsymbol{k'}}^{\gamma\gamma'}=(1+\hat\kv\cdot\hat\kv')/2$.  Considering the short-range disorder  
characterized by an effective strength of $|U_{\kv -\boldsymbol{k'}}|\approx n_dV_0^2$ ($n_d$ is the disorder density and $V_0\delta(\bf r)$ is the disorder potential), the linearized collision integral becomes
\begin{equation}
\begin{split}
\Iv_D(\kv ,\gamma) =\frac{2\pi g v  n_dV_0^2}{\hbar}\sum_{\kv',\gamma}\left(\frac{\partial f^0_{\kv ,\gamma '}}{\partial \epsilon_{\kv ,\gamma} } \right)(1+\hat\kv\cdot\hat\kv')(\hat\kv-\hat\kv')\cdot(\pv_n+\beta\tilde{\epsilon}_{\kv,\gamma}\pv_s)\delta(\epsilon_{\kv ,\gamma }-\epsilon_{\kv ',\gamma})
\end{split}
\end{equation}
From this we easily construct the moments of the disorder collision integral
\begin{equation}\label{EqID1}
\begin{split}
I_D^{11} &=\frac{2\pi g^2 v ^2 n_dV_0^2}{\hbar}\sum_{\kv}\sum_{\kv',\gamma}\left(\frac{\partial f^0_{\kv ,\gamma '}}{\partial \epsilon_{\kv ,\gamma} } \right)[1-(\hat\kv\cdot\hat\kv')^2]\delta(\epsilon_{\kv ,\gamma }-\epsilon_{\kv ',\gamma})\\
I_D^{12} &=I_D^{21}=\frac{2\pi g^2 v ^2 n_dV_0^2}{\hbar}\sum_{\kv}\sum_{\kv',\gamma}\beta\tilde{\epsilon}_{\kv,\gamma}\left(\frac{\partial f^0_{\kv ,\gamma '}}{\partial \epsilon_{\kv ,\gamma} } \right)[1-(\hat\kv\cdot\hat\kv')^2]\delta(\epsilon_{\kv ,\gamma }-\epsilon_{\kv ',\gamma})\\
I_D^{22} &=\frac{2\pi g^2v ^2 n_dV_0^2}{\hbar}\sum_{\kv}\sum_{\kv',\gamma}(\beta\tilde{\epsilon}_{\kv,\gamma})^2\left(\frac{\partial f^0_{\kv ,\gamma '}}{\partial \epsilon_{\kv ,\gamma} } \right)[1-(\hat\kv\cdot\hat\kv')^2]\delta(\epsilon_{\kv ,\gamma }-\epsilon_{\kv ',\gamma}).
\end{split}
\end{equation}
Using  $x=\beta\epsilon_{\kv,+}$ and $\bar\mu=\beta\mu$ the integrals (\ref{EqID1}) can be simplified to 
\begin{equation}
\begin{split}
I_D^{11} & =\alpha\int_{0}^\infty x^2~dx\left[\frac{e^{x-\bar{\mu}}}{ (1+e^{x-\bar{\mu}})^2} +\frac{e^{-x-\bar{\mu}}}{ (1+e^{-x-\bar{\mu}})^2 }\right],\\
I_D^{12} =I_D^{21} &  =\alpha\int_{0}^\infty x^2~dx\left[\frac{(x-\bar{\mu})e^{x-\bar{\mu}}}{ (1+e^{x-\bar{\mu}})^2}-\frac{(x+\bar{\mu})e^{-x-\bar{\mu}}}{ (1+e^{-x-\bar{\mu}})^2 }\right],\\
I_D^{22} &  =\alpha\int_{0}^\infty x^2~dx\left[\frac{(x-\bar{\mu})^2e^{x-\bar{\mu}}}{ (1+e^{x-\bar{\mu}})^2}+\frac{(x+\bar{\mu})^2e^{-x-\bar{\mu}}}{ (1+e^{-x-\bar{\mu}})^2 }\right],
\end{split}
\end{equation}
 where $\alpha=\lambda/(\pi\hbar^3\beta^2)$ with $\lambda=n_dV_0^2/(\hbar v )^2$. We will present our results in terms of the dimensionless disorder strength $\lambda$.  The final analytical solutions are
\begin{equation}\label{EqID}
I_D^{11}  =
\alpha\left[\bar{\mu}^2+\frac{\pi^2}{3}\right],~~~
I_D^{12}=I_D^{21}  =\alpha\frac{2\bar{\mu}\pi^2}{3},~~~
I_D^{22} =\alpha
\left[ \frac{\bar{\mu}^2\pi^2}{3}+\frac{7\pi^4}{15}\right].
\end{equation}

\end{widetext}

\end{document}